\documentclass[onecolumn,english]{IEEEtran}
\usepackage[T1]{fontenc}
\usepackage{babel}
\usepackage{color,amssymb}

\usepackage{amsthm}
\usepackage{amsmath}
\usepackage{amssymb}
\usepackage[unicode=true,
 bookmarks=true,bookmarksnumbered=true,bookmarksopen=true,bookmarksopenlevel=1,
 breaklinks=false,pdfborder={0 0 0},backref=false,colorlinks=false]
 {hyperref}
\hypersetup{pdftitle={Your Title},
 pdfauthor={Your Name},
 pdfpagelayout=OneColumn, pdfnewwindow=true, pdfstartview=XYZ, plainpages=false}

\makeatletter

\providecommand{\tabularnewline}{\\}

\theoremstyle{plain}
\newtheorem{thm}{\protect\theoremname}
\theoremstyle{remark}
\newtheorem{claim}[thm]{\protect\claimname}
\theoremstyle{plain}
\newtheorem{lem}[thm]{\protect\lemmaname}

\ifCLASSOPTIONcompsoc
\usepackage[caption=false,font=normalsize,labelfont=sf,textfont=sf]{subfig}
\else
\usepackage[caption=false,font=footnotesize]{subfig}
\fi

\makeatother

\providecommand{\claimname}{Claim}
\providecommand{\lemmaname}{Lemma}
\providecommand{\theoremname}{Theorem}

\begin{document}

\title{SUPER: Sparse signals with Unknown Phases Efficiently Recovered}

\author{\IEEEauthorblockN{Sheng Cai, \and Mayank Bakshi, \and Sidharth Jaggi, \and Minghua Chen}\\
 \texttt{\small \{cs010,mayank,jaggi,minghua\}@ie.cuhk.edu.hk}{\small }\\
{\small{} \IEEEauthorblockA{The Chinese University of Hong Kong,
Hong Kong SAR China}}}
\maketitle

\begin{abstract}
Suppose ${\bf x}$ is any exactly $k$-sparse vector in $\mathbb{C}^{n}$. We present a class of phase measurement matrix $A$ in $\mathbb{C}^{m\times n}$, and a corresponding algorithm, called SUPER, that can resolve ${\bf x}$ up to a global phase from intensity measurements $|A{\bf x}|$ with high probability over $A$. Here $|A{\bf x}|$ is a vector of component-wise magnitudes of $A{\bf x}$. The SUPER algorithm is the first to simultaneously have the following properties: (a) it requires only ${\cal O}(k)$ (order-optimal) measurements, (b) the computational complexity of decoding is ${\cal O}(k\log k)$ (near order-optimal) arithmetic operations.
\end{abstract}

\section{Introduction}
\noindent {\bf \underline{Phase Retrieval:}} In many applications, it's difficult to measure the phase information of the underlying signal. Instead, we recover the signal by its intensity measurements. For instance, in X-ray crystallography, optics \cite{mil:90} and image reconstruction for astronomy \cite{DauF:87}, signal/image is reconstructed from the intensity measurements of its Fourier transform.

Let $A \in \mathbb{C}^{m \times n}$ be used to denote the {\it phase measurement matrix}, and ${\bf x}\in\mathbb{C}^{n}$ be used to denote the unknown underlying signal. Instead of {\it linear} measurements of the form $y = A{\bf x}$ as in the {\it compressive sensing} literature (see, for instance,~\cite{CRT:06}) in the {\it phase retrieval problem} we have $m$ {\it non-linear} {\it intensity measurements} of the form $b_i = |<A_i,{\bf x}>|$. Here the index $i$ is an integer in $\{1,\ldots,m\}$ (or [m] for short), $A_{i}$ is the $i$-th row of phase measurement matrix $A$, $<\cdot> $
is the inner product and $\left|\cdot\right|$ is the absolute value.

Problems of this kind have been studied over the last decades. A good survey of some of the algorithms via non-convex process can be found in~\cite{Fie:82,Fie:13}. Recently, two convex optimization methods, PhaseLift \cite{CPA:13} and PhaseCut \cite{Waldm:13}, have been proposed by Cand\`{e}s {\it et al.} and Waldspurger {\it et al.}. PhaseLift is inspired by finding the low-rank matrix (specifically for the phase retrieval problem, rank-one matrices) by minimizing the trace norm (SDP) \cite{CanR:2012}. PhaseLift is able to reconstruct ${\bf x}$ with ${\cal O}(n\log n)$ intensity measurements by solving semidefinite programming with high probability. The $A_i$'s are independently sampled on the unit sphere of $\mathbb{C}^{n}$. Later, it's shown that the number of intensity measurements can be improved to ${\cal O}(n)$ where $A_i$'s are independently and identically distributed with the uniform distribution on the sphere of radius $\sqrt n$, or the complex normal distribution \cite{CanL:13}. PhaseCut is inspired by solving max-cut problem via SDP. The decoding complexity for both PhaseLift and PhaseCut is ${\cal O}\left(n^3\right)$, which is still computationally costly when $n$ is large.

Besides SDP-based approach, more computationally efficient algorithms are proposed such as \cite{AleBFM:12}, \cite{NetJS:13}. For instance, in \cite{NetJS:13}, the number of intensity measurements required is ${\cal O}\left(n\log^{3} n\right)$. However, the decoding complexity is ${\cal O}\left(n^{2}\log^{3} n\right)$ which is less than that of SDP-based approach.

\noindent {\bf \underline{Compressive Phase Retrieval:}}
Suppose ${\bf x}$ is ``sparse'', {\it i.e.}, the number of non-zero components of ${\bf x}$ is at most $k$, which is much less than the length $n$ of ${\bf x}$. This assumption is not uncommon in many applications like X-ray crystallography. Then, given $A$ and $b$,  the goal of {\it compressive phrase retrieval} is to reconstruct ${\bf x}$ as $\hat{{\bf x}}$, where $\hat{{\bf x}}$ equals ${\bf x}$ up to a global phase. That is, $\hat{{\bf x}} = {\bf x}e^{\iota\Theta}$ for some arbitrary fixed $\Theta \in [0,2\pi)$. Here $\iota$ denotes the positive square root of $-1$.   The reason we allow this degeneracy in $\hat{{\bf x}}$, up to a global phase factor, is that all such $\hat{\bf x}$'s result in the same measurement vector under intensity measurements. If $\hat{{\bf x}}$ does indeed equal ${\bf x}$ up to a global phase, then we denote this ``equality'' as $\hat{{\bf x}} \hat{=} {\bf x}$.

%
%

It is shown that $4k-1$ intensity measurements suffice to uniquely reconstruct ${\bf x}$ in \cite{HenY:13} (for ${\bf x}\in \mathbb{R}^{n}$) and \cite{MehV:13} (for ${\bf x}\in \mathbb{C}^{n}$). However, no efficient algorithms is given. The $\ell_{1}$-regularized PhaseLift method is introduced in the compressive phase retrieval problem in~\cite{OhlYDS:11}. In \cite{LiV:12}, it is shown that if the number of Gaussian intensity measurements is ${\cal O}\left(k^{2}\log n\right)$, ${\bf x}$ can be correctly reconstructed via $\ell_{1}$-regularized PhaseLift.

The works in \cite{SheBE:13} and the works by Jaganathan {\it et al.} \cite{JanOH:12,JanOH:13,JanOH:121} study the case when the phase measurement matrix is a Fourier transform matrix. In \cite{SamAMYB:13}, it is explained that SDP-based methods can reconstruct ${\bf x}$ with sparsity up to $o\left(\sqrt n\right)$. In \cite{JanOH:13}, the algorithm based on reweighted $\ell_1$-minimization with ${\cal O}\left(k^{2}\log n\right)$ phaseless Fourier measurements is proposed to go beyond this bottleneck. When the phase measurement matrix is allowed to be designed, a matrix ensemble and a corresponding combinatorial algorithm is proposed in \cite{JanOH:13} such that ${\bf x}$ is correctly reconstructed with ${\cal O}(k\log n)$ intensity measurements in ${\cal O}(kn\log n)$ time.

To our best knowledge, in the literature, there is no construction of a measurement matrix $A$ and a corresponding reconstruction algorithm that correctly reconstructs ${\bf x}$ with an order-optimal number of measurements and with near-optimal decoding complexity simultaneously.


\begin{table}[tbh]
\centering{}%
\begin{tabular}{c|l}
\hline
Notation & Definition\tabularnewline
\hline
\hline
${\bf x}$ & Length-$n$ signal over $\mathbb{C}$ with sparsity $k$\tabularnewline
\hline
$A$ & Dimension-$n\times m$ phase measurement matrix over $\mathbb{C}$.\tabularnewline
\hline
$b$ & Length-$m$ Intensity measurement vector over $\mathbb{R}^+$.\tabularnewline
\hline
$A_{i}$ & The $i$-th row of phase measurement matrix $A$ for all $\forall i\in[m]$.\tabularnewline
\hline
$b_{i}$ & $b_{i}=\left|\left\langle A_{i},\ {\bf x}\right\rangle \right|$,
the $i$-th intensity measurement $\forall i\in[m]$.\tabularnewline
\hline
$k$ & $k=\left\Vert {\bf x}\right\Vert _{0}$, the number of non-zero components
(sparsity) of ${\bf x}$.\tabularnewline
\hline
\end{tabular}\caption{Table of notation for the model}
\end{table}

\subsection{Our Contribution}

In this work, we describe a randomized design of the phase measurement matrix $A$ and
a corresponding decoding algorithm achieving the following guarantees:
\begin{thm}\label{thm:main}
(Main theorem) There exists a measurement ensemble $\{A\}$ and a corresponding decoding algorithm for compressive phase
retrieval with the following performance:\end{thm}
\begin{enumerate}
\item For every ${\bf x}\in\mathbb{C}^{n}$, with probability $1-o(1)$
over the randomized design of $A$, the algorithm exactly reconstructs
${\bf x}$ up to a global phase;
\item The number of measurements $m={\cal O}(k)$;
\item The decoding complexity is ${\cal O}(k\log k)$.
\end{enumerate}


The rest of this paper is organized as follows. We first present the high-level overview of our algorithm in Section \ref{sec:intuition}. In Section \ref{sec:graph}, we introduce the graphs used for measurement structure. Section \ref{sec:measurement} and Section \ref{sec:parameter} contain actual measurement design. Section \ref{sec:reconstruction} and Section \ref{sec:performance} discuss the reconstruction algorithm and the performance of it. Section \ref{sec:conclude} concludes this paper.

\section{Overview/High-level Intuition}
\label{sec:intuition}
Our SUPER algorithm is non-adaptive. There are three phases%
\footnote{All the measurements are designed before the decoding process, so
it is still non-adaptive.%
} in our decoding algorithm. In the first phase (called seeding phase),
we are able to recover the magnitudes and relative phases of constant
fraction of non-zero components of ${\bf x}$. In the second phase (called geometric-decay phase), there
are ${\cal O}(\log(\log k))$ stages. In each stage, we recovery
the magnitudes and relative phases of constant fraction of unresolved
non-zero components of ${\bf x}$. In the third phase (called cleaning-up phase), the remaining
${\cal O}(k/\log k)$ unresolved non-zero components are decoded.

\subsection{Pieces of the puzzle}

We first define some useful terminology.

\noindent {\underline{{\it Singletons:}}}

If a measurement $b_i$ involves only a single non-zero component of ${\bf x}$, then we say that such a measurement is a {\it singleton}.\footnote{We borrow this terminology (of singletons, doubletons, multitons, etc) from the compressive sensing work of Pawar {\it et al}~\cite{PawR:12}.} Singletons are important since they can be used to pin down the magnitude (though not the phase) of components of ${\bf x}$. There are several challenges, however. One lies in even identifying whether a measurement is a singleton or not. The second lies in identifying which of the ${\bf x}$ components being measured in $b_i$ corresponds to the singleton. The third is to be able to do all this blindingly fast, in fact in {\it constant} time (independent of $n$ and $k$!). Each of these challenges can be handled by using ideas from the our prior work on compressive sensing~\cite{SHO-FA}. For details, see Sections \ref{sec:measurement} and~\ref{sec:reconstruction} below.

\noindent {\underline{{\it Doubletons:}}}

Similarly, if a measurement $b_i$ involves exactly two non-zero components of ${\bf x}$, then we say that such a measurement is a {\it doubleton}. Doubletons, especially doubletons measuring two non-zero components of ${\bf x}$ which have already been measured by singletons (we call such doubletons {\it resolvable doubletons}), are useful since they can be used to deduce the relative phases of the two non-zero components of ${\bf x}$. For example, if one is given the magnitudes $|x_i|$, $|x_j|$, and $|x_i + x_j|$, then one can determine the angle $\theta$  between the phases of the complex numbers $x_i$ and $x_j$ (up to degeneracy of sign of $\theta$). In fact, even this degeneracy can be resolved by an additional judiciously chosen measurement. Similar challenges to those mentioned above vis-a-vis singletons (identifying whether or not a measurement is a doubleton/resolvable doubleton, identifying which components of ${\bf x}$ it corresponds to, and doing so in constant time) also hold for doubletons. See Sections \ref{sec:measurement} and~\ref{sec:reconstruction} for details.

\noindent {\underline{{\it Mutual resolvability:}}}

We say our decoding algorithm has thus far {\it mutually resolved} two non-zero components $x_i$ and $x_j$ of ${\bf x}$ if the magnitudes of both $x_i$ and $x_j$ have been deduced, and also the relative phase between $x_i$ and $x_j$ has been deduced (for instance via resolvable doubleton measurements roughly described above). Note that mutual resolvability is an equivalence relation -- it is reflexive, symmetric and transitive. Note therefore that if $x_i$ and $x_{i'}$ have been mutually resolved, it is not necessary that they even are involved in the same measurement; it is sufficient that $x_i$ and $x_{i'}$ are part of a chain of non-zero components of ${\bf x}$ that are pairwise mutually resolved. Finally, we note that as our decoding algorithm progresses, if it is successful, in fact {\it all} the non-zero components of ${\bf x}$ are eventually mutually resolved. Hence this property of mutual resolvability is perhaps most interesting in the intermediate stages of our decoding algorithm.

\noindent {\underline{{\it Giant component:}}}

We say that a subset of the non-zero components of ${\bf x}$ form a giant component if it is the largest subset satisfying the two properties:

\begin{itemize}

\item The subset is of size linear in $k$.

\item Any pair of components in the subset have been mutually resolved (thus far) by the decoding algorithm.

\end{itemize}

Non-zero components of ${\bf x}$ that have not (yet) been mutually resolved with respect to an element of the giant component by the decoding algorithm are said to be unresolved.

Essentially, our algorithm proceeds by iteratively enlarging the giant component until it engorges all the non-zero components of ${\bf x}$.

\noindent {\underline{{\it Resolvable multiton:}}}

We say that a measurement $b_i$ is a resolvable multiton if it is the case that exactly one (say $x_i$) of the non-zero components of ${\bf x}$ involved in the measurement $b_i$ is outside the giant component, and at least one of non-zero components of ${\bf x}$ is inside the giant component. Such measurements are useful since, in the latter parts of our algorithm, there are not enough resolvable doubletons. By carefully choosing the parameters of the algorithm, one can guarantee that a constant fraction of measurements are resolvable mutitons.

Judiciously designed measurements (see Section~\ref{sec:measurement}) enable one to mutually resolve the component $x_i$ that is outside the giant component, with the components of ${\bf x}$ inside the giant component, by solving a quadratic equation. Care is indeed required in choosing the measurements since the amplitude measurement process is inherently non-linear, and there may not be a ``clean'' manner to mutually resolve $x_i$ via arbitrary measurements -- indeed the design of such a measurement process is also one of the intellectual contributions we wish to highlight in this work. We call this process ``cancelling out'' the already resolved components of ${\bf x}$.

\subsection{Putting the pieces together}

\noindent {\underline{{\it Seeding phase:}}}

In the first phase, called the {\it seeding phase}, there are ${\cal O}(k)$ ``sparse'' measurements (each measurement involves, in expectation, ${\cal O}(n/k)$ components of ${\bf x}$). We demonstrate that by first examining the measurements corresponding to this phase, the decoding algorithm is already able to decode a constant fraction (say $1/2$)\footnote{Here, $1/2$ is arbitrarily chosen to simplify the presentation of intuition. The actual fraction of resolved non-zero components in the seeding phase is different from $1/2$. See Section \ref{sec:performance} for details. Here, the parameter $1/2$ for the geometric-decay phase in this section is due to the same reason.} of the components of ${\bf x}$ up to a global phase. The algorithm is able to do this since we are able to show that a ``significant'' fraction of measurements are singletons and resolvable doubletons. Standard results in percolation theory \cite{Bol:01} then lead one to conclude that the number of non-zero nodes that are mutually resolvable is linear in $k$, {\it i.e.}, that there is a giant component. Hence this phase is called the ``seeding'' phase, since the giant component forms the nucleus on which the remainder of the algorithm builds upon.

Prior work (\cite{JanOH:13}) closest to our work here comprises essentially only of the seeding phase, but with ${\cal O}(k\log(k))$\footnote{The combinatorial algorithm in \cite{JanOH:13} can be modified to have ${\cal O}(k\log(k))$ measurement with error probability ${\cal O}(1/poly(k))$ instead of $1/n$ in the paper. Also, based on our reconstruction algorithm, the decoding complexity can be reduced to ${\cal O}(k\log k)$.} measurements. The reason that prior work needs this many measurements is essentially due to what happens at the tail end of a ``coupon collection'' process~\cite{MotR:95} (wherein one has to collect at least one copy of each of $k$ coupons by sampling with replacement) -- when most of the coupons have already been collected/the giant component is of size close to $k$, then the growth rate slows down. Specifically, this is because the fraction of resolvable doubletons decays slowly to zero, and an additional multiplicative factor of $\log(k)$ measurements is required so as to ensure the giant component subsumes all non-zero components of ${\bf x}$.

The key technique used in our work, then, is to segue to a different sampling process outlined below, and using resolvable multitons rather than doubletons. The challenge is to make the numbers work -- unlike~\cite{JanOH:13}, not only do we require only ${\cal O}(k)$ measurements, but we also require our decoding complexity to be ${\cal O}(k\log(k))$.

\noindent {\underline{{\it Geometric-decay phase:}}}

This phase itself comprises of ${\cal O}(\log(\log(k)))$ separate stages. Each stage has half the number of measurements compared to the previous stage, but measurements in each stage are twice as ``dense'' as the measurements in the previous stage. So, for instance, if in the first stage of the geometric-decay phase, there are say $ck$ measurements, with each measurement involving $n/k$ components of ${\bf x}$, then in the second stage of the geometric-decay phase, there are $ck/2$ measurements, but each measurement involves $2n/k$ components of ${\bf x}$.

There are two reasons for this choice of parameters. Firstly, with such a geometric decay in the number of measurements in each stage, the overall number of measurements in the geometric-decay phase is still ${\cal O}(k)$. Secondly, we show that with the geometric increase in the density of measurements, a significant fraction of measurements in each stage lead to resolvable multitons, and use this to show that the number of unresolved components decays geometrically.

The reason we run the geometric-decay phase for only ${\cal O}(\log(\log(k)))$ stages is also two-fold. Firstly, after that many stages, with the number of unresolved components halving at every stage, the number of unresolved components of ${\bf x}$ is, in expectation, ${\cal O}(k/\log(k))$. Hence the concentration inequalities (which depend on the number of unresolved components) we use to control the probabilities of error get progressively weaker (though they still result in good concentration at the last stage of the geometric-decay phase). Secondly, and more importantly, the number of non-zero components in each resolvable multiton increases geometrically as the number of stages increases. This has implications for the time-complexity of the decoding algorithm, since the time-complexity depends directly on the number of non-zero components in each measurement that need to be ``cancelled out''. By terminating the geometric-decay phase after ${\cal O}(\log(\log(k)))$ stages ensures that, in expectation, the number of such ``cancellations'' is at most ${\cal O}(\log(k))$, and hence the overall time-complexity of the algorithm scales as ${\cal O}(k\cdot\log(k))$.

\noindent {\underline{{\it Cleaning-up phase:}}}

Finally, we segue to what we call the ``cleaning-up'' phase. As noted above, after the geometric-decay phase the number of unresolved components of ${\bf x}$ is, in expectation, $k' \triangleq {\cal O}(k/\log(k))$. To fit our budget of ${\cal O}(k)$ measurements, and ${\cal O}(k\log(k))$ decoding time, we now segue to using ``coupon collection'' as a primitive. This may be viewed as restarting the seeding (first) phase, but with different parameters. In particular, the problem dimension has now been significantly reduced (since there are now only $k'$ unresolved components of ${\bf x}$). Therefore we can now afford to pay the coupon collection penalty that we avoided in the seeding phase by moving to the geometric-decay phase.

Specifically, in this cleaning-up phase we take ${\cal O}(k'\log(k'))$ measurements so as to resolve the remaining $k'$ unresolved components of ${\bf x}$. Note that ${\cal O}(k'\log(k'))$ scales as ${\cal O}(k)$. Each measurement we take has the same density as the measurements in the last stage of the geometric decay phase, and hence the time-complexity of resolving measurements also scales in the same manner. However, since there are many more measurements than in the last stage of the geometric-decay phase, by standard arguments corresponding to the coupon collection problem we are able to argue that for each unresolved component of ${\bf x}$ there is at least one resolvable multiton that helps resolve it.

\subsection{Summary of the overview}
\label{sec:graph}
As the above discussion outlines, to make the numbers work ({\it i.e.}, to ensure ${\cal O}(k)$ number of measurements and ${\cal O}(k\log(k))$ time-complexity), one has to delicately choose the parameters of the measurement ensemble. Our analysis indicates that having a phase in which the sparsity actually geometrically increases, at least for a while, significantly improves performance. To take advantage of this, however, we have to carefully design the measurements, so that one can resolve unresolved components of ${\bf x}$ via judiciously designed non-linear measurements. In this work we have not attempted to optimize the constant factors -- we expect further constant-factor improvements are possible via further careful tuning.

\section{Graph properties}

We construct a series of bipartite graphs with some desirable properties
outlined in this section. We then use the structure of the bipartite
graphs to generate our measurement matrix $A$ in Section \ref{sec:measurement} and
design the corresponding reconstruction algorithm in Section \ref{sec:reconstruction}.
Each left nodes of a bipartite graph represents a component of ${\bf x}$
and each right node represents a set of intensity measurements.

\subsection{Seeding Phase}

The properties of the bipartite graph, ${\cal G}_{I}$, in the first
phase are as follows:
\begin{enumerate}
\item There are $n$ left nodes and $ck$ right nodes, where $c$ is a constant.
\item Each edge in ${\cal G}_{I}$ appears with probability $1/k$. For
each right node, the degree, in expectation, is $n/k$.
\item For each edge in ${\cal G}_{I}$, it is assigned different weights
which are discussed in the measurement design (See Section \ref{sec:measurement}).
\item Many singleton nodes: Singleton nodes are right nodes which involves
exactly one non-zero component of ${\bf x}$. Singleton nodes help to recover
the magnitude of non-zero component. See Section \ref{sec:performance} for details.
\item Many resolvable doubleton nodes: Doubleton nodes are right nodes which
involve exactly two non-zero components of ${\bf x}$. Resolvable doubletons
are the doubletons which involve exactly two non-zero components whose
magnitudes are recovered by singleton nodes. See Section \ref{sec:performance} for details.
\end{enumerate}
Another graph ${\cal H}$ is implied by ${\cal G}_{I}$. Each vertex
in ${\cal H}$ represents a non-zero component of ${\bf x}$ and there
is an edge in ${\cal H}$ if and only if two left nodes involved are
mutually resolved by a resolvable doubleton node. The property of ${\cal H}$
is as follows:
\begin{enumerate}
\item ${\cal H}$ has a giant connected component: The connected component, ${\cal H}'$
contains a constant fraction of nodes in ${\cal H}$. This property
is formally stated in Section \ref{sec:performance}.
\end{enumerate}

\subsection{Geometric-decay phase}

There are ${\cal O}(\log\log k)$ separate bipartite graphs/stages
in this phase.

The properties of the $l$-th bipartite graph, ${\cal G}_{II,l}$ ($l=1,2,\ldots,L={\cal O}(\log\log k)$),
are as follows:
\begin{enumerate}
\item There are $n$ left nodes and $cf_{II,l-1}k$ right nodes, where $f_{II,l-1}$
is the expected fraction of unresolved non-zero components of ${\bf x}$
after the ($l-1$)-th stage of decoding process in the second phase. $f_{II,0}=f_{I}$ is the expected fraction of unresolved non-zero components after seeding phase. The $0$-th stage of geometric-decay phase is seeding-phase. The
value of $f_{II,l}$ is discussed in Section \ref{sec:parameter}.
\item Each edges in ${\cal G}_{II,l}$ appears with probability $1/\left(f_{II,l-1}k\right)$.
\item For each edge in ${\cal G}_{II,l}$, it is assigned different weights
which are discussed in the measurement design.
\item Many resolvable multiton nodes: The resolvable multiton nodes are right nodes
which involve exactly one unresolved non-zero component of ${\bf x}$ and
at least one of the resolved non-zero components. Each resolvable multiton
node helps to recover both the magnitude and the relative phase of
the corresponding unresolved non-zero component via ``Cancelling
out'' process (See Section \ref{sec:reconstruction}).
\end{enumerate}
For a newly resolved non-zero component, the corresponding node in
${\cal H}$ is appended to the giant connected component, ${\cal H}'$. In expectation,
there are $\left(f_{II,l-1}-f_{II,l}\right)k$ non-zero components
decoded in the $l$-th stage of decoding. We show in Section \ref{sec:performance}
that we are able to reconstruct a constant fraction of undecoded non-zero
components with high probability at each stage. After ${\cal O}(\log\log k)$
stages, there are ${\cal O}(k/\log k)$ unresolved non-zero components of
${\bf x}$ left.

\subsection{Cleaning-up phase}

The properties of the bipartite graph, ${\cal G}_{III}$, in the last
phase are as follows:
\begin{enumerate}
\item There are $n$ left nodes and $c\left(k/\log k\right)\log\left(k/\log k\right)={\cal O}(k)$
right nodes.
\item Each edges in ${\cal G}_{III}$ appears with probability $\log k/k$.
\item For each edge in ${\cal G}_{III}$, it is assigned different weights
which are discussed in the measurement design.
\item Many resolvable multiton nodes.
\end{enumerate}
In this stage, all the resolved non-zero components of size ${\cal O}(k/\log k)$ are finally recovered using resolvable multiton nodes by ``Cancelling out'' process and a Coupon Collection argument.

\begin{center}
\begin{table}[tbh]
\begin{centering}
\begin{tabular}{c|l}
\hline
Notation & Definition\tabularnewline
\hline
\hline
${\cal G}_{I}$ & The bipartite graph used in the seeding phase with $n$ left nodes
and $ck$ right nodes.\tabularnewline
 & Each edge appears with probability $1/k$.\tabularnewline
\hline
${\cal H}$ & Implied graph by ${\cal G}_{I}$.\tabularnewline
\hline
${\cal H}'$ & Connected component of ${\cal H}$.\tabularnewline
\hline
${\cal G}_{II,l}$ & The $l$-th bipartite graph used in the $l$-th stage in geometric-decay
phase with $n$ left nodes\tabularnewline
 &  and $cf_{II,l-1}k$ right nodes for $l\in[L]$. Each
edge appears with probability $1/f_{II,l-1}k$.\tabularnewline
\hline
$f_{I}$ & The expected fraction of unresolved non-zero components of ${\bf x}$
after the seeding phase.\tabularnewline
\hline
$f_{II,l}$ & The expected fraction of unresolved non-zero components of ${\bf x}$
after the $l$-th stage\tabularnewline
 & of the geometric-decay phase. Let $f_{II,0}=f_{I}$.\tabularnewline
\hline
${\cal G}_{III}$ & The bipartite graph used in the cleaning-up phase with $n$ left nodes
and \tabularnewline
 & $c\left(k/\log k\right)\log\left(k/\log k\right)$ right nodes. Each
edge appears with probability $\log k/k$.\tabularnewline
\hline
\end{tabular}
\par\end{centering}

\caption{Table of notation used in the design of bipartite graphs}

\end{table}

\par\end{center}
\section{Measurement Design}
\label{sec:measurement}
For a bipartite graph ${\cal G}$ (${\cal G}$ is one of the ${\cal G}_{I}$,
${\cal G}_{II,l}$'s and ${\cal G}_{III}$), there are $n$ nodes
on the left and $m_{{\cal G}}'$ nodes on the right . $A({\cal G})'$
is the dimension-$m_{{\cal G}}'\times n$ adjacent matrix of ${\cal G}$
where the entry at $i$-th row and $j$-th column equals to $1$ if
and only if $i$-th right node connects to the $j$-th left node for
$j\in[n]$ and $i\in\left[m_{{\cal G}}'\right]$. The dimension-$m_{{\cal G}}\times n$
phase measurement matrix $A({\cal G})$ is designed based on $A({\cal G})'$ where
$m_{{\cal G}}=5m_{{\cal G}}'$. By appending all the matrix $A({\cal G})$
sequentially, we get the actual $m\times n$ measurement matrix $A$
where $m=\Sigma_{{\cal G}}m_{{\cal G}}$. For $i$-th row $A({\cal G})'_{i}$
of $A({\cal G})'$, a set of rows (of size $5$) of $A({\cal G})$
are designed for $i\in\left[m_{{\cal G}}'\right]$. If the $j$-th
entry of $A({\cal G})'_{i}$ is zero, then corresponding set of entries
of $A({\cal G})$ are all zero for all $j\in[n]$. In the following
measurement matrix design, we design the entries corresponding to
non-zero entries in $A({\cal G})'$. See Section \ref{sec:reconstruction} for how these measurements are used for decoding.
\begin{enumerate}
\item {\underline {Trigonometric entries}}: The $j$-th entries of the $(5i-4)$-th and $(5i-3)$-th
rows of $A({\cal G})$ are denoted by $a_{i,j}^{({\cal G},1)}$ and $a_{i,j}^{({\cal G},2)}$
. The values are set as follows:
\begin{eqnarray*}
a_{i,j}^{({\cal G},1)} & = & \cos\left(\frac{j\pi}{2n}\right)\\
a_{i,j}^{({\cal G},2)} & = & \iota\sin\left(\frac{j\pi}{2n}\right),
\end{eqnarray*}
where $\iota$ denotes the positive square root of $-1$ and $\pi/2n$ can be treated as the unit phase of the entry design.
In particular, the phase $j\pi/2n$ will be critical for our algorithm.
The first two entries are used in singleton node identification and
``cancelling out'' process of resolvable multiton node.
\item {\underline {Structured unit complex entries}}: The $j$-th entry of the $(5i-2)$-th row of
$A({\cal G})$ is denoted by $a_{i,j}^{({\cal G},3)}$ . The value is
set as follows:
\begin{eqnarray*}
a_{i,j}^{({\cal G},3)} & = & \exp\left(\iota\frac{j\pi}{2n}\right).
\end{eqnarray*}
This type
of measurement will be used only in ``cancelling out''
process of resolvable multiton node.
\item {\underline {Unit entries}}: The $j$-th entry of the $(5i-1)$-th row of $A$
is denoted by $a_{i,j}^{({\cal G},4)}$.
The value is set to be $1$. This measurement is used in resolvable
doubleton identification and ``cancelling out'' process
of resolvable multiton node.
\item {\underline {Random unit complex entries}}: The $j$-th entry of the $5i$-th row of $A$
is denoted by $a_{i,j}^{({\cal G},5)}$ used as verification. The value
is set as follows:
\[
a_{i,(j)}^{({\cal G},5)}=\exp(\iota\phi_{i,j}),
\]
where $\iota$ denotes the positive square root of $-1$ and $\phi_{i,j}$
is chosen uniformly at random from $[0,\ \pi/2]$. This measurement
is used in resolve the degeneracy when resolvable multiton and resolvable
doubleton are used for decoding. Also, it helps to verify our identification
and estimation of magnitude and relative phase.
\end{enumerate}

\begin{center}
\begin{table}[tbh]
\begin{centering}
\begin{tabular}{c|l}
\hline
Notation & Definition\tabularnewline
\hline
\hline
$m'_{{\cal G}}$ & The number of right nodes for the bipartite graph ${\cal G}$. ${\cal G}$
is one of ${\cal G}_{I}$, ${\cal G}_{II,l}$ for $l\in[L]$, and
${\cal G}_{III}$.\tabularnewline
\hline
$A({\cal G})'$ & The dimension-$m'_{{\cal G}}\times n$ adjacent matrix of ${\cal G}$.\tabularnewline
\hline
$A({\cal G})'_{i}$ & The $i$-th row of matrix $A({\cal G})'$ for $i\in\left[m'_{{\cal G}}\right]$.\tabularnewline
\hline
$A({\cal G})$ & The dimension-$m_{{\cal G}}\times n$ measurement matrix generated
by $A({\cal G})'$. Here $m_{{\cal G}}=5m'_{{\cal G}}$.\tabularnewline
\hline
$A$ & The dimension-$m\times n$ phase measurement matrix generated by all
$A({\cal G})$'s. Here, $m=\Sigma_{{\cal G}}m_{{\cal G}}$.\tabularnewline
\hline
$a_{i,j}^{\left({\cal G},q\right)}$ & The $j$-th entry of the $\left[5\left(i-1\right)+q\right]$-th the
rows of $A({\cal G})$. Here, $i\in \left[m_{\cal G}\right]$, $j\in [n]$, and $q\in[5]$.\tabularnewline
\hline
\end{tabular}
\par\end{centering}

\caption{Table of notation for measurements design}

\end{table}

\par\end{center}
\section{Reconstruction Algorithm}
\label{sec:reconstruction}

Let $b_{i}^{\left({\cal G},q\right)}$ denote the $\left[5\left(i-1\right)+q\right]$-th measurement generated by $A({\cal G})$. Here, ${\cal G}$ is one of the ${\cal G}_{I}$,
${\cal G}_{II,l}$'s and ${\cal G}_{III}$, $i\in \left[m_{\cal G}\right]$, and $q\in[5]$.

\subsection{Seeding phase}

\subsubsection{Overview}
\begin{enumerate}
\item Preprocessing: Each right node is attached to a list to record its
neighbors (left nodes) in the decoding process.
\item Magnitude Recovery and Doubleton Identification: Check every right node to see whether it's singleton node
or not. If yes, we locate the corresponding non-zero component and
measure the magnitude of it. After decoding the non-zero component
(only the magnitude), each list of its neighbors' (right nodes') is
inserted the location of the decoded non-zero component if the length
of the list is no larger than one. For the list whose length is $3$
after insertion, it will be discarded and won't be considered in the
following iteration since it definitely is not a doubleton. So far
we get the potential resolvable doubletons. Later, we use the verification
measurement to find the actual resolvable doubletons. The reason why we
need the verification step is that the potential resolvable doubletons
may involve other non-zero components which have not been resolved yet.

In this step, we decode the magnitudes of constant fraction of all
the non-zero components and locate these non-zero components. We also identify
the potential resolvable doubletons by checking whether its list is
of length $2$ and the actual resolvable doubletons by verification
measurement.

\item Relative Phase Recovery: For each resolvable doubleton, it's used to resolve the phase
between the two non-zero components whose locations lie in the neighbor
list.

Breadth first search (BFS) or Depth first search (DFS) \cite{Tar:72} algorithm would
guide us to explore the connected components in graph ${\cal H}$ efficiently.
We only care about the largest connected component, ${\cal H}'$.
After this step, any pair of nodes in ${\cal H}'$ are mutually resolved.
\end{enumerate}

\subsubsection{The formal description of reconstruction algorithm}
\begin{enumerate}
\item Initialization: We initialize by setting the signal estimate vector
$\hat{{\bf x}}$ to all-zeros vector $0^{n}$. Each right node $i\in\left[m_{{\cal G}_{I}}'\right]$
attaches an empty neighbor list ${\cal N}(i)$. let ${\cal D}$ denote a list of the resolvable
doubletons. Initially, ${\cal D}$ is empty. Set $i=1$.
\item Singleton Identification, Magnitude Recovery and Doubleton Identification:

\begin{enumerate}
\item Compute the ratio of Trigonometric measurements:
\[
s_{i}=\frac{\arctan\left(\frac{b_{i}^{\left({\cal G}_{I},2\right)}}{\iota b_{i}^{({\cal G}_{I},1)}}\right)}{\frac{\pi}{2n}}.
\]

\begin{enumerate}
\item Check if $s_{i}$ is an integer. If so, we tentatively identifies
that $i$ is a singleton, $s_{i}$-th entry of ${\bf x}$ is non-zero
and
\[
\left|\hat{x}_{s_{i}}\right|=\begin{cases}
\frac{b_{i}^{\left({\cal G}_{I},1\right)}}{a_{i,s_{i}}^{\left({\cal G}_{I},1\right)}} & \mbox{if }a_{i,s_{i}}^{\left({\cal G}_{I},1\right)}\neq0\\
\iota \frac{b_{i}^{\left({\cal G}_{I},2\right)}}{a_{i,s_{i}}^{\left({\cal G}_{I},2\right)}} & \mbox{if }a_{i,s_{i}}^{\left({\cal G}_{I},2\right)}\neq0.
\end{cases}
\]

\end{enumerate}
\item We verify our estimate from the previous step. If $\left|\hat{x}_{s_{i}}\right|\neq\left|b_{i}^{\left({\cal G}_{I},5\right)}\right|$,
the verification fails. We increment $i$ by $1$ and go back to step a) to start
a new iteration. If verification passes, we do the following steps:

\begin{enumerate}
\item $s_{i}$ is appended to the neighbor lists of all its neighbors. For
$i\in\left[m_{{\cal G}_{I}}'\right]$, it is no longer considered
in the later process if $\left|{\cal N}(i)\right|\geq3$ since in
the next step we only care about doubleton whose neighbor list size
equals $2$.
\item Increment $i$ by $1$ and go back to step a) to start a new iteration.
\end{enumerate}
\item For each $i$ whose neighbor list is of size $2$,
 it is appended to the resolvable doubleton list ${\cal D}$ where ${\cal N}(i)[1]$
and ${\cal N}(i)[2]$ are the two indices of non-zero components whose
magnitudes have been recovered.
\end{enumerate}
\item Relative Phase Recovery:

\begin{enumerate}
\item Compute connected component of ${\cal H}$: Breadth first search or
depth first search for adjacent list representation of ${\cal H}$ is applied in this step.
For each $i\in{\cal D}$, the elements in ${\cal N}(i)$ tell which
two vertices in ${\cal H}$ are connected. BFS or DFS outputs
connected components of graph ${\cal H}$. We run the BFS or DFS,
for each edge in ${\cal H}$, with additional steps b), c), and d) stated below:
\item Law of Cosine: Suppose $i$'s two neighbors are denoted by ${\cal N}(i)[1]$
and ${\cal N}(i)[2]$. The fourth measurement is
used to derive the phase between ${\cal N}(i)[1]$-th and ${\cal N}(i)[2]$-th
components of ${\bf x}$, $\theta=\left|\theta_{{\cal N}(i)[1]}-\theta_{{\cal N}(i)[2]}\right|$,
by Law of Cosine%
\footnote{Given the lengths of two complex number $A$ and $B$, we can deduce
the phase between $A$ and $B$, $\Delta$, by Law of Cosine if we also know the
length of $A+B$. To be more explicit, $-\cos\Delta=\frac{|A|^{2}+|B|^{2}-|A+B|^{2}}{2|A||B|}$. %
}.
\item The verification measurement helps to resolve the degeneracy of sign
of $\theta$ ({\it i.e.}, whether $\theta$ or $-\theta$ is the actual
phase difference we are interested in.) by checking whether
\begin{eqnarray*}
\left|\left|\hat{{\bf x}}_{{\cal N}(i)[1]}\right|\exp\left(\iota\phi_{i,{\cal N}(i)[1]}\right)+\left|\hat{{\bf x}}_{{\cal N}(i)[2]}\right|\exp\left(\iota\phi_{i,{\cal N}(i)[2]}+\iota\theta\right)\right|=\left|b_{i}^{\left({\cal G}_{I},5\right)}\right|
\end{eqnarray*}

or
\begin{eqnarray*}
\left|\left|\hat{{\bf x}}_{{\cal N}(i)[1]}\right|\exp\left(\iota\phi_{i,{\cal N}(i)[1]}\right)+\left|\hat{{\bf x}}_{{\cal N}(i)[2]}\right|\exp\left(\iota\phi_{i,{\cal N}(i)[2]}-\iota\theta\right)\right|=\left|b_{i}^{\left({\cal G}_{I},5\right)}\right|.
\end{eqnarray*}

If neither of the above equations holds, then $i$ is not a resolvable doubleton.

\item For the first node in a connected component, its phase is set to be
zero.
\item When the BFS or DFS terminates, we can find the largest connected
component of ${\cal H}$, ${\cal H}'$. For all the node pairs in
${\cal H}'$, they are mutually resolved.
\end{enumerate}
\end{enumerate}

\subsection{Geometric-decay and Cleaning-up phases}
\begin{claim}
(``Cancelling out'' Process) For a bipartite graph ${\cal G}$
in geometric-decay phase or cleaning-up phase, if a right node $i$
is a resolvable multiton node, it involves exactly one (unknown) undecoded
non-zero component, ${\bf x}_{j}$, and at least one (known) resolved
non-zero components. Then, we are able to find the location of ${\bf x}_{j}$,
$j$, and resolve ${\bf x}_{j}$ (both magnitude and relative phase).\end{claim}
\begin{IEEEproof}
We will use four measurements in the ``cancelling out''
process,

\begin{eqnarray*}
b_{i}^{\left({\cal G},1\right)} & = & \left|A+{\bf x}_{j}\cos\left(\frac{j\pi}{2n}\right)\right|\\
b_{i}^{\left({\cal G},2\right)} & = & \left|B+{\bf x}_{j}\iota\sin\left(\frac{j\pi}{2n}\right)\right|\\
b_{i}^{\left({\cal G},3\right)} & = & \left|C+{\bf x}_{j}\exp\left(\iota\frac{j\pi}{2n}\right)\right|\\
b_{i}^{\left({\cal G},5\right)} & = & \left|D+{\bf x}_{j}\exp\left(\iota\phi_{i,j}\right)\right|,
\end{eqnarray*}

where $A$, $B$, $C$, and $D$ are calculated from the decoded non-zero
components which connect to right node $i$ in ${\cal G}$.

We find that by the measurements design
\begin{eqnarray*}
A+B & = & C
\end{eqnarray*}

and

\begin{eqnarray*}
{\bf x}_{j}\cos\left(\frac{j\pi}{2n}\right)+{\bf x}_{j}\iota\sin\left(\frac{j\pi}{2n}\right) & = & {\bf x}_{j}\exp\left(\iota\frac{j\pi}{2n}\right).
\end{eqnarray*}

Let

\begin{eqnarray*}
\frac{j\pi}{2n} & = & \alpha\\
A+{\bf x}_{j}\cos\left(\frac{j\pi}{2n}\right) & = & U\\
B+{\bf x}_{j}\iota\sin\left(\frac{j\pi}{2n}\right) & = & V\\
C+{\bf x}_{j}\exp\left(\iota\frac{j\pi}{2n}\right) & = & W,
\end{eqnarray*}

we have
\begin{eqnarray*}
b_{i}^{\left({\cal G},1\right)} & = & \left|U\right|\\
b_{i}^{\left({\cal G},2\right)} & = & \left|V\right|\\
b_{i}^{\left({\cal G},3\right)} & = & \left|W\right|\\
 & = & \left|U+V\right|.
\end{eqnarray*}

\noindent {\underline {Finding the relation between $U$ and $V$}:}

We know that
\begin{eqnarray*}
U & = & V\times\frac{b_{i}^{\left({\cal G},1\right)}}{b_{i}^{\left({\cal G},2\right)}}\exp\left(\iota\psi\right),
\end{eqnarray*}

or

\begin{eqnarray*}
U & = & V\times\frac{b_{i}^{\left({\cal G},1\right)}}{b_{i}^{\left({\cal G},2\right)}}\exp\left(-\iota\psi\right),
\end{eqnarray*}

where $\psi$ is the phase between $U$ and $V$ and $\cos\psi=\frac{\left|U\right|^{2}+\left|V\right|^{2}-\left|U+V\right|^{2}}{2\left|U\right|\left|V\right|}$.

\noindent {\underline {Finding the relation between ${\bf x}$ and $\alpha$:}}

For simplicity, we only consider the case that

\begin{eqnarray*}
U & = & V\times\frac{b_{i}^{\left({\cal G},1\right)}}{b_{i}^{\left({\cal G},2\right)}}\exp\left(\iota\psi\right)\\
 & \triangleq & V\times M.
\end{eqnarray*}

So,

\begin{eqnarray*}
A+{\bf x}_{j}\cos\alpha & = & \left[B+{\bf x}_{j}\iota\sin\alpha\right]M.
\end{eqnarray*}

We have
\begin{eqnarray*}
{\bf x}_{j} & = & \frac{BM-A}{\cos\alpha-\iota M\sin\alpha}.
\end{eqnarray*}

\noindent {\underline {Solving $\cos^{2}\alpha$ by quadratic equation:}}

Replacing ${\bf x}_{j}$in
\[
b_{i}^{\left({\cal G},1\right)}=\left|U\right|,
\]

we know that
\begin{eqnarray*}
b_{i}^{\left({\cal G},1\right)} & = & \left|A+\frac{BM-A}{\cos\alpha-\iota M\sin\alpha}\cos\alpha\right|\\
 & = & \left|\frac{BM\cos\alpha-\iota AM\sin\alpha}{\cos\alpha-\iota M\sin\alpha}\right|\\
 & = & \left|\frac{B\cos\alpha-\iota A\sin\alpha}{\cos\alpha-\iota M\sin\alpha}\right|\left|M\right|\\
 & = & \left|\frac{B\cos\alpha-\iota A\sin\alpha}{\cos\alpha-\iota M\sin\alpha}\right|\frac{b_{i}^{\left({\cal G},1\right)}}{b_{i}^{\left({\cal G},2\right)}}.
\end{eqnarray*}

So,
\begin{eqnarray*}
b_{i}^{\left({\cal G},2\right)}\left|\cos\alpha-\iota M\sin\alpha\right| & = & \left|B\cos\alpha-\iota A\sin\alpha\right|.
\end{eqnarray*}

Let
\begin{eqnarray*}
A & = & A_{1}+\iota A_{2}\\
B & = & B_{1}+\iota B_{2}\\
M & = & M_{1}+\iota M_{2},
\end{eqnarray*}

where $A_1$, $A_2$, $B_1$, $B_2$, $M_1$, and $M_2$ are real numbers. We have
\begin{eqnarray*}
b_{i}^{\left({\cal G},2\right)}\left|\cos\alpha-\iota\left(M_{1}+\iota M_{2}\right)\sin\alpha\right|\\
=\left|\left(B_{1}+\iota B_{2}\right)\cos\alpha-\iota\left(A_{1}+\iota A_{2}\right)\sin\alpha\right|.
\end{eqnarray*}

Squaring both sides, we get

\begin{eqnarray*}
\left[b_{i}^{\left({\cal G},2\right)}\right]^{2}\left[\left(\cos\alpha+M_{2}\sin\alpha\right)^{2}+\left(M_{1}\sin\alpha\right)^{2}\right]\\
=\left(B_{1}\cos\alpha+A_{2}\sin\alpha\right)^{2}+\left(B_{2}\cos\alpha-A_{1}\sin\alpha\right)^{2}.
\end{eqnarray*}

After reorganizing the above equation, we have

\begin{eqnarray*}
\left(\left[b_{i}^{\left({\cal G},2\right)}\right]^{2}-\left|B\right|^{2}\right)\cos^{2}\alpha+\left(\left[b_{i}^{\left({\cal G},1\right)}\right]^{2}-\left|A\right|^{2}\right)\sin^{2}\alpha\\
=2\cos\alpha\sin\alpha\left(A_{2}B_{1}-A_{1}B_{2}-2\left[b_{i}^{\left({\cal G},2\right)}\right]^{2}M_{2}\right).
\end{eqnarray*}

Let
\begin{eqnarray*}
P & = & \left[b_{i}^{\left({\cal G},2\right)}\right]^{2}-\left|B\right|^{2}\\
Q & = & \left[b_{i}^{\left({\cal G},2\right)}\right]^{2}-\left|A\right|^{2}\\
R & = & A_{2}B_{1}-A_{1}B_{2}-2\left[b_{i}^{\left({\cal G},2\right)}\right]^{2}M_{2}\\
S & = & \cos^{2}\alpha
\end{eqnarray*}

and square both sides, we have
\begin{eqnarray*}
\left[PS+Q(1-S)\right]^{2} & = & 4R^{2}S(1-S).
\end{eqnarray*}

After reorganizing the above equation, we get

\begin{eqnarray*}
\left(P^{2}+Q^{2}-2PQ+4R^{2}\right)S^{2}\\
+\left(2PQ-2Q^{2}-4R^{2}\right)S+Q^{2} & = & 0.
\end{eqnarray*}

We are able to solve $S$ (quadratic equation) in constant time and
similarly for the case that $U=V\times\frac{b_{i}^{\left({\cal G},1\right)}}{b_{i}^{\left({\cal G},2\right)}}\exp\left(-\iota\psi\right)$.

\noindent {\underline {Resolving the degeneracy via random unit complex measurements:}}

After deriving the value of $S=\cos^{2}\alpha$, we can get the constant
($4$) possible value of $j$ and ${\bf x}_{j}$ (both magnitude and
the relative phase in ${\cal H}'$) pairs.

Last, we check which pairs of solution that satisfies the following equation
to resolve the degeneracy
\begin{eqnarray*}
b_{i}^{\left({\cal G},5\right)} & = & \left|D+{\bf x}_{j}\exp\left(\iota\phi_{i,j}\right)\right|.
\end{eqnarray*}

\end{IEEEproof}
Note that if ``cancelling out'' fails ({\it i.e.,} none of the pairs of $j$ and ${\bf x}_j$ satisfies the last equation in the proof), then $i$ is not a resolvable multiton.
In each stage at geometric-decay phase and cleaning-up phase, we go through all the right nodes, find resolvable multitons and use them to recover unresolved non-zero components by the ``cancelling out'' process. For a newly resolved component of ${\bf x}$, the corresponding node in ${\cal H}$ is appended to ${\cal H}'$. In the end, the size of the node set of ${\cal H}'$ should be $k$.

\begin{center}
\begin{table}[tbh]
\begin{centering}
\begin{tabular}{c|l}
\hline
Notation & Definition\tabularnewline
\hline
\hline
$b_{i}^{\left({\cal G},q\right)}$ & The $\left[5\left(i-1\right)+q\right]$-th intensity measurement generated
by measurement matrix $A({\cal G})$. Here, $i\in\left[m_{{\cal G}}\right]$,
and $q\in[5]$. \tabularnewline
\hline
${\cal D}$ & Resolvable doubleton list used in the seeding phase.\tabularnewline
\hline
${\cal S}$ & Singleton List. used in the seeding phase.\tabularnewline
\hline
${\cal N}(i)$ & The neighbor list for $i$-th node in ${\cal G}_{I}$ for $i\in\left[m_{{\cal G}_{I}}\right]$.\tabularnewline
\hline
\end{tabular}
\par\end{centering}

\caption{Table of notation for measurements design}

\end{table}

\par\end{center}

\section{Parameters Design}
\label{sec:parameter}

All the parameters designed in this section are calculated based on expectation. The actual performance of our algorithm will be discussed in Section \ref{sec:performance}.

\subsection{Seeding phase}

\subsubsection{Magnitude Recovery by singletons}
\begin{itemize}
\item The probability of a right node being a singleton node:
\end{itemize}
\begin{eqnarray*}
P_{S} & = & {k \choose 1}\frac{1}{k}\left(1-\frac{1}{k}\right)^{k-1}\\
 & = & \left(1-\frac{1}{k}\right)^{k-1}\\
 & \doteq & e^{-1}.
\end{eqnarray*}

\begin{itemize}
\item The expected number of singletons is $=ck\times P_{S}\doteq e^{-1}ck$.
\end{itemize}
\begin{itemize}
\item The expected number of different non-zero components whose magnitudes are recovered:\end{itemize}
\begin{lem}
\label{lem:couponcollection}
(Generalized coupon collection) Given $V$ different coupons and $V\log\frac{V}{V-U}$ picks with repetition ($U<V$), the expected number of
different coupons picked is $U$ for $V\rightarrow+\infty$. With probability at least $1-2\exp\left(-\frac{2\epsilon^{2}(V-U)U}{V}\right)$, the number of different coupons picked is between
$(1-\epsilon)U$ and $(1+\epsilon)U$ for any $\epsilon >0$.\end{lem}

By Lemma \ref{lem:couponcollection} (let $V=k$ and $V\log\frac{V}{V-U}=ck\times P_{S}$),
we know that the expected number of non-zero components of $x$ whose magnitudes
are recovered is $k\left(1-e^{-cP_{S}}\right)$.

\subsubsection{Relative Phase Recovery by resolvable doubletons}
\begin{itemize}
\item The probability of a right node being doubleton:
\end{itemize}
\begin{eqnarray*}
P_{D} & = & {k \choose 2}\left(\frac{1}{k}\right)^{2}\left(1-\frac{1}{k}\right)^{k-2}\\
 & = & \frac{1}{2}\left(1-\frac{1}{k}\right)^{k-1}\\
 & \doteq & \frac{e^{-1}}{2}.
\end{eqnarray*}

\begin{itemize}
\item The expected number of doubletons is $=ck\times P_{D}\doteq e^{-1}ck/2$.
\end{itemize}
\begin{itemize}
\item The expected number of resolvable doubletons:
\end{itemize}

Note that only the doubleton which involves two non-zero components whose
magnitudes have been recovered is useful to recover the relative
phase.

\begin{eqnarray*}
\mbox{\# resolvable doubletons} & = & \frac{{k\left(1-e^{-cP_{S}}\right) \choose 2}}{{k \choose 2}}\times ckP_{D}\\
 & \doteq & \frac{\left(1-e^{-cP_{S}}\right)^{2}cke^{-1}}{2}.
\end{eqnarray*}

\begin{itemize}
\item The expected number of different pairs of components whose relative phase is recovered by resolvable doubletons:
\end{itemize}

By Lemma \ref{lem:couponcollection}, given $k\left(1-e^{-cP_{S}}\right)$ nodes and $\left(1-e^{-cP_{S}}\right)^{2}ckP_D$ edges with repetition in ${\cal H}$,
there are

\[
\left(1+{\cal O}(1/k)\right)\left(1-e^{-cP_{S}}\right)^{2}ckP_D
\]

distinct edges.
\subsubsection{The giant connected components}

\begin{thm}
\label{thm:giantcomponent}\cite{Bol:01} For a random graph ${\cal G}_{N,M}$ with
$N$ nodes and $M$ edges chosen at random among the ${N \choose 2}$
possible edges. Let ${\cal Z}_{N,M}$ denote the size of the greatest
component of ${\cal G}_{N,M}$. If $r=2M/N>1$, we have for any $\epsilon>0$
\begin{eqnarray*}
\Pr\left(\left|\frac{{\cal Z}_{N,M}}{N}-\beta\right|<\epsilon\right) & = & 1-{\cal O}\left(\frac{1}{\epsilon^{2}N}\right),
\end{eqnarray*}
where $\beta$ is the unique solution to $\beta+\exp(-\beta r)=1$.

\end{thm}
We need to find the size of giant connected component of a random
graph with $k\left(1-e^{-cP_{S}}\right)$ nodes and $\left(1-e^{-cP_{S}}\right)^{2}ckP_D$
edges (with repetition) and therefore $\left(1+{\cal O}(1/k)\right)\left(1-e^{-cP_{S}}\right)^{2}ckP_D$ distinct edges (implied by Lemma \ref{lem:couponcollection}). Let's say the size is $\left(1-f_{I}\right)k$ where $f_{I}$
is the function of $c$.

By Theorem \ref{thm:giantcomponent}, when $2\left(1-e^{-cP_{S}}\right)cP_D>1$,
the giant connected component exists (this inequality holds when constant
$c$ is large enough) and the size of the giant component is $\left(1-f_{I}\right)k=\beta_{c}k\left(1-e^{-cP_{S}}\right)$
where $\beta_{c}$ is the unique solution to $\beta+\exp\left[-\beta\cdot2\left(1-e^{-cP_{S}}\right)cP_D\right]=1$.

\subsection{Geometric-decay phase:}

Let $f_{II,l}$ denote the expected fraction of unresolved non-zero components
after the $l$-th stages in this phase. Let $f_I=f_{II,0}$.

\subsubsection{Stage $l+1$ ($0\leq l\leq L-1$)}
\begin{itemize}
\item The probability that a right node being a resolvable multiton:
\end{itemize}
\begin{eqnarray*}
P_{M}^{(II,l+1)} & = & {f_{II,l}k \choose 1}\frac{1}{f_{II,l}k}\left(1-\frac{1}{f_{II,l}k}\right)^{f_{II,l}k-1}\left[1-\left(1-\frac{1}{f_{II,l}k}\right)^{\left(1-f_{II,l}\right)k}\right]\\
 & = & \left(1-\frac{1}{f_{II,l}k}\right)^{f_{II,l}k-1}-\left(1-\frac{1}{f_{II,l}k}\right)^{k-1}\\
 & \doteq & e^{-1}-e^{-\frac{1}{f_{II,l}}}.
\end{eqnarray*}

\begin{itemize}
\item The expected number of resolvable multitons is $cf_{II,l}kP_{M}^{(II,l+1)}$.
\end{itemize}
\begin{itemize}
\item The expected number of non-zero components which are resolved (both
magnitude and phase):
\end{itemize}

By Lemma \ref{lem:couponcollection}, let
\[
f_{II,l}k\log\frac{f_{II,l}k}{f_{II}k-(f_{II,l}-f_{II,l+1})k}=cf_{II,l}kP_{M}^{(II,l+1)},
\]

we know that $f_{II,l}-f_{II,l+1}=f_{II,l}\left(1-e^{-cP_{M}^{(II,l+1)}}\right)$.

Therefore, $f_{II,l+1}=e^{-cP_{M}^{(II,l+1)}}f_{II,l}$. We can compute the value of $f_{II,l}$ recursively.

Note that $P_{M}^{(II,l)}$ increases as $l$ increases. So, $P_{M}^{(II,l)}$ is bounded by $e^{-1}-e^{-\frac{1}{f_{I}}}$ ($l=0$) and $e^{-1}$ ($l=+\infty$).

\subsubsection{End of this phase}

There are ${\cal O}(\log\log k)$ stages in the geometric-decay phase. We already show that in each step we expect to recover constant fraction
of remaining unresolved non-zero components. In the end of this phase, the number of unresolved non-zero components is ${\cal O}(k/\log k$).

\subsection{Cleaning-up phase}

Recall that, in this phase, each edges appears with probability $\log k/k$ and there are $c\left(k/\log k\right)\log\left(k/\log k\right)={\cal O}(k)$ right nodes in ${\cal G}_{III}$.

%
%
%

\section{Performance of algorithm (Proof of Main Theorem)}
\label{sec:performance}

\noindent {\underline {\bf Number of measurements}}:

In Section \ref{sec:parameter}, we already know that $f_{II,l+1}=e^{-cP_{M}^{(II,l+1)}}f_{II,l}$. And $P_{M}^{(II,l)}$ increases as $l$ increases.
\begin{eqnarray*}
f_{II,l+1} & = & \exp\left[-cP_{M}^{(II,l+1)}\right]f_{II,l}\\
 & = & \exp\left[-c\Sigma_{t=1}^{l+1}P_{M}^{\left(II,t\right)}\right]f_{II,0}\\
 & = & \exp\left[-c\Sigma_{t=1}^{l+1}P_{M}^{\left(II,t\right)}\right]f_{I}\\
 & \leq & \exp\left[-c(l+1)P_{M}^{\left(II,1\right)}\right]f_{I}.
\end{eqnarray*}

Then, the total number of measurements in three phases is
\begin{eqnarray*}
\underbrace{ck}_{Seeding}+\underbrace{\text{\ensuremath{\Sigma}}_{l=1}^{L}cf_{II,l-1}k}_{geometric-decay}+\underbrace{c(k/\log k)\log(k/\log k)}_{cleaning-up} & = & {\cal O}(k)+\left(\text{\ensuremath{\Sigma}}_{l=1}^{L}cf_{II,l-1}\right)k+{\cal O}(k)\\
 & \leq & {\cal O}(k)+\left(\Sigma_{l=1}^{L}\exp\left[-clP_{M}^{\left(II,1\right)}\right]\right)f_{I}ck\\
 & = & {\cal O}(k).
\end{eqnarray*}

\noindent {\underline {\bf Decoding complexity}}:

Almost all the operations take constant time except for DFS in the seeding phase and ``Cancelling out'' process in the geometric-decay and cleaning-up  phases.

For DFS, the time complexity is linear in the size of node set and edge set. Since there are $k$ nodes and ${\cal O}(k)$ edges involved in the seeding phase, the time complexity is ${\cal O}(k)$.

For ``Cancelling out'' process, the time complexity is dominated by  calculating the value of $A$, $B$, $C$ and $D$ (See Section \ref{sec:reconstruction}). And the complexity depends on the number of resolved non-zero components which corresponds to the resolvable multiton.

In the later stage/phase, more non-zero components are associated with a measurement. Since the number of measurements is ${\cal O}(k)$, it suffices to show that each measurement involves at most ${\cal O}(\log k)$ non-zero components (even if they are unresolved) in the cleaning-up phase with probability at least $1-o(1/k)$.

Let $NZ$ be the number of non-zero components involved in a measurement
in cleaning-up phase. By Chernoff bound, for any $\epsilon_{NZ}\geq 0$, we have

\begin{eqnarray*}
\Pr\left[NZ\geq\left(1+\epsilon_{NZ}\right)k\cdot\log k/k\right] & \leq & \exp\left(-\frac{\epsilon_{NZ}^{2}}{2+\epsilon_{NZ}^{2}}k\cdot\log k/k\right).
\end{eqnarray*}

Therefore,

\begin{eqnarray*}
\Pr\left[NZ={\cal O}(\log k)\right] & \geq & 1-{\cal O}(1/poly(k)).
\end{eqnarray*}

Thus, we know that the decoding complexity is at most ${\cal O}(k\cdot NZ)={\cal O}(k\log k)$ with probability at least $1-k\cdot {\cal O}(1/poly(k))=1-o(1)$ by Union bound.

\noindent {\underline {\bf Correctness}}:

The actual performance of our algorithm is slightly
different from expectation. Here, ``slightly'' means that the actual
number of resolved non-zero components in each phase/stage deviates from the expected value
but it can be concentrated around expectation with high probability.

Let $g_{I}$ denote the actual fraction of unresolved non-zero components
after seeding phase. Let $g_{II,l}$ denote the actual fraction of unresolved
non-zero components after the $l$-th stage in geometric-decay phase. Let $g_{II,0}=g_I$.

Just recall the properties of the bipartite graphs for measurement design. In the seeding phase, each edge appears with probability $1/k$ and there
are $ck$ right nodes. In the geometric-decay phase, each edge appears with probability
$1/f_{II,l}k$ and there are $cf_{II,l}k$ right nodes in ($l+1$)-th
step for $l\geq 0$. In the cleaning-up phase, each edge appears with probability $\log k/k$
and there are $c(k/\log k)\log(k/\log k)$ right nodes.

\subsection{Seeding Phase}

\subsubsection{Magnitude recovery}
\begin{itemize}
\item By Chernoff bound, the probability that the number of singletons is
larger than $\left(1+\epsilon_{S}\right)ck\times P_{S}$ or smaller
than $\left(1-\epsilon_{S}\right)ck\times P_{S}$ is less than $2e^{-\left(\epsilon_{S}\right)^{2}ckP_{S}/2}={\cal O}\left(\exp\left(-\epsilon_{S}^{2}k\right)\right)$ for any $\epsilon_{S}>0$.
\item By Lemma \ref{lem:couponcollection} and Union bound, we know that, for any $\epsilon_{DS}>0$, the number of different
non-zero components whose magnitudes are recovered is between $\left(1-\epsilon_{DS}\right)\left[1-e^{-\left(1-\epsilon_{S}\right)cP_{S}}\right]k$
and $\left(1+\epsilon_{DS}\right)\left[1-e^{-\left(1+\epsilon_{S}\right)cP_{S}}\right]k$
with probability
\[
1-{\cal O}\left[\exp\left(-\epsilon_{DS}^{2}k\right)+\exp\left(-\epsilon_{S}^{2}k\right)\right].
\]
Note that
\[
\left(1+\beta_{S}\right)\left[1-e^{-\left(1+\epsilon_{S}\right)cP_{S}}\right]k
\]
and
\[
\left(1-\beta_{S}\right)\left[1-e^{-\left(1-\epsilon_{S}\right)cP_{S}}\right]k
\]
scale as
\[
\left(1-e^{-cP_{S}}\right)\left[1+(\epsilon_{DS}+\epsilon_{S})+o(\epsilon_{DS}+\epsilon_{S})\right]k
\]
and
\[
\left(1-e^{-cP_{S}}\right)\left[1-(\epsilon_{DS}+\epsilon_{S})-o(\epsilon_{DS}+\epsilon_{S})\right]k,
\]
respectively.
\end{itemize}

\subsubsection{Relative phase recovery}
\begin{itemize}
\item By Chernoff bound, the probability that the number of doubletons is
larger than $\left(1+\epsilon_{D}\right)ck\times P_{D}$ or smaller
than $\left(1-\epsilon_{D}\right)ck\times P_{D}$ is less than $2e^{-\left(\epsilon_{D}\right)^{2}ckP_{D}/2}={\cal O}\left(\exp\left(-\epsilon_{D}^{2}k\right)\right)$ for any $\epsilon_{D}>0$.
\item The resolvable doubletons
\end{itemize}
\begin{eqnarray*}
\mbox{\# resolvable doubletons} & \leq & \frac{{\left(1+\epsilon_{DS}\right)\left[1-e^{-\left(1+\epsilon_{S}\right)cP_{S}}\right]k \choose 2}}{{k \choose 2}}\\
 &  & \times\left(1+\epsilon_{D}\right)\left(1+\epsilon_{RD}\right)ckP_{D}
\end{eqnarray*}

and

\begin{eqnarray*}
\mbox{\# resolvable doubletons} & \geq & \frac{{\left(1-\epsilon_{DS}\right)\left[1-e^{-\left(1-\epsilon_{S}\right)cP_{S}}\right]k \choose 2}}{{k \choose 2}}\\
 &  & \times\left(1-\epsilon_{D}\right)\left(1-\epsilon_{RD}\right)ckP_{D}
\end{eqnarray*}

with probability
\[
1-{\cal O}\left[\exp\left(-\epsilon_{DS}^{2}k\right)+\exp\left(-\epsilon_{S}^{2}k\right)+\exp\left(-\epsilon_{D}^{2}k\right)+\exp\left(-\epsilon_{RD}^{2}k\right)\right],
\]
for any $\epsilon_{RD}>0$, by Chernoff bound and Union bound. Again, the upper bound and the
lower bound on the number of resolvable doubletons scale as

\[
\left[1+{\cal O}\left(\epsilon_{DS}+\epsilon_{S}+\epsilon_{D}+\epsilon_{RD}\right)\right]\left(1-e^{-cP_{S}}\right)^{2}ckP_{D}
\]

and

\[
\left[1-{\cal O}\left(\epsilon_{DS}+\epsilon_{S}+\epsilon_{D}+\epsilon_{RD}\right)\right]\left(1-e^{-cP_{S}}\right)^{2}ckP_{D},
\]
 respectively.
\begin{itemize}
\item Number of distinct edges in giant component: By Lemma \ref{lem:couponcollection}
and Union bound, for any $\epsilon_{DRD}>0$, with probability
\[
1-{\cal O}\left[\exp\left(-\epsilon_{DS}^{2}k\right)+\exp\left(-\epsilon_{S}^{2}k\right)+\exp\left(-\epsilon_{D}^{2}k\right)+\exp\left(-\epsilon_{RD}^{2}k\right)+\exp\left(-\epsilon_{DRD}^{2}k\right)\right],
\]
 the number of pairs of relative phase resolved by all the resolvable
doubletons will be bounded by
\[
\left(1+\epsilon_{DRD}\right)\left[1+{\cal O}\left(\epsilon_{DS}+\epsilon_{S}+\epsilon_{D}+\epsilon_{RD}\right)/k\right]\left[1+{\cal O}\left(\epsilon_{DS}+\epsilon_{S}+\epsilon_{D}+\epsilon_{RD}\right)\right]\left(1-e^{-cP_{S}}\right)^{2}ckP_{D}
\]
 and
\[
\left(1-\epsilon_{DRD}\right)\left[1-{\cal O}\left(\epsilon_{DS}+\epsilon_{S}+\epsilon_{D}+\epsilon_{RD}\right)/k\right]\left[1-{\cal O}\left(\epsilon_{DS}+\epsilon_{S}+\epsilon_{D}+\epsilon_{RD}\right)\right]\left(1-e^{-cP_{S}}\right)^{2}ckP_{D}
\]
 which scale as
\[
\left[1+{\cal O}\left(\epsilon_{DS}+\epsilon_{S}+\epsilon_{D}+\epsilon_{RD}+\epsilon_{DRD}\right)\right]\left(1-e^{-cP_{S}}\right)^{2}ckP_{D}
\]
and
\[
\left[1-{\cal O}\left(\epsilon_{DS}+\epsilon_{S}+\epsilon_{D}+\epsilon_{RD}+\epsilon_{DRD}\right)\right]\left(1-e^{-cP_{S}}\right)^{2}ckP_{D}.
\]

\end{itemize}

\subsubsection{The giant connected component}

Let $N^{+}$and $N^{-}$ be the upper bound and lower bound on the
number of nodes in giant component.

Let $M^{+}$ and $M^{-}$ be the upper bound and lower bound on the
number of edges in giant component. Then, $r^{+}=2M^{+}/N^{-}$ is
the upper bound on twice the size of edges over size of nodes and
the $r^{-}=2M^{-}/N^{+}$ is the lower bound. $\beta_{c}^{+}$ and
$\beta_{c}^{-}$ are the solution to the equation $\beta+\exp\left(-\beta r^{+}\right)=1$
and $\beta+\exp\left(-\beta r^{-}\right)=1$.

We know that
\begin{eqnarray*}
r^{+} & = & 2M^{+}/N^{-}\\
 & = & 2\left[1+{\cal O}\left(\epsilon_{DS}+\epsilon_{S}+\epsilon_{D}+\epsilon_{RD}+\epsilon_{DRD}\right)\right]\left(1-e^{-cP_{S}}\right)^{2}ckP_{D}\\
 &  & /\left(1-e^{-cP_{S}}\right)\left[1-{\cal O}(\epsilon_{DS}+\epsilon_{S})\right]k\\
 & = & 2\left[1+{\cal O}\left(\epsilon_{DS}+\epsilon_{S}+\epsilon_{D}+\epsilon_{RD}+\epsilon_{DRD}\right)\right]\left(1-e^{-cP_{S}}\right)cP_{D}
\end{eqnarray*}

and

\begin{eqnarray*}
r^{-} & = & 2M^{-}/N^{+}\\
 & = & 2\left[1-{\cal O}\left(\epsilon_{DS}+\epsilon_{S}+\epsilon_{D}+\epsilon_{RD}+\epsilon_{DRD}\right)\right]\left(1-e^{-cP_{S}}\right)^{2}ckP_{D}\\
 &  & /\left(1-e^{-cP_{S}}\right)\left[1+{\cal O}(\epsilon_{DS}+\epsilon_{S})\right]k\\
 & = & 2\left[1+{\cal O}\left(\epsilon_{DS}+\epsilon_{S}+\epsilon_{D}+\epsilon_{RD}+\epsilon_{DRD}\right)\right]\left(1-e^{-cP_{S}}\right)cP_{D}.
\end{eqnarray*}

Since $r=-\log(1-\beta)/\beta$, $dr/d\beta=\left(-\log(1-\beta)/\beta\right)'_{\beta=\beta_{c}}$
is a constant.

By Theorem \ref{thm:giantcomponent},
\begin{eqnarray*}
\Pr\left(\left|\frac{{\cal Z}_{N,M}}{N}-\beta_{c}\right|\leq\epsilon_{GC}\right) & = & {\cal O}\left(\frac{1}{\epsilon_{GC}^{2}k}\right),
\end{eqnarray*}

for any $\epsilon_{GC}>0$.

Therefore,
\[
\beta_{c}^{+}=\left[1+{\cal O}\left(\epsilon_{DS}+\epsilon_{S}+\epsilon_{D}+\epsilon_{RD}+\epsilon_{DRD}\right)\right]\left(\beta_{c}+\epsilon_{GC}\right)
\]

and

\[
\beta_{c}^{-}=\left[1-{\cal O}\left(\epsilon_{DS}+\epsilon_{S}+\epsilon_{D}+\epsilon_{RD}+\epsilon_{DRD}\right)\right]\left(\beta_{c}-\epsilon_{GC}\right).
\]

The upper bound on the size of giant component is

\begin{eqnarray*}
N^{+}\beta_{c}^{+} & = & \left(1-e^{-cP_{S}}\right)\left[1+{\cal O}(\epsilon_{DS}+\epsilon_{S})\right]k\left[1+{\cal O}\left(\epsilon_{DS}+\epsilon_{S}+\epsilon_{D}+\epsilon_{RD}+\epsilon_{DRD}\right)\right]\left(\beta_{c}+\epsilon_{GC}\right)\\
 & = & \left[1+{\cal O}\left(\epsilon_{DS}+\epsilon_{S}+\epsilon_{D}+\epsilon_{RD}+\epsilon_{DRD}+\epsilon_{GC}\right)\right]\beta_{c}\left(1-e^{-cP_{S}}\right)k
\end{eqnarray*}

and the lower bound on the size of giant component is
\begin{eqnarray*}
N^{-}\beta_{c}^{-} & = & \left[1-{\cal O}\left(\epsilon_{DS}+\epsilon_{S}+\epsilon_{D}+\epsilon_{RD}+\epsilon_{DRD}+\epsilon_{GC}\right)\right]\beta_{c}\left(1-e^{-cP_{S}}\right)k,
\end{eqnarray*}

with probability

\[
1-{\cal O}\left[\exp\left(-\epsilon_{DS}^{2}k\right)+\exp\left(-\epsilon_{S}^{2}k\right)+\exp\left(-\epsilon_{D}^{2}k\right)+\exp\left(-\epsilon_{RD}^{2}k\right)+\exp\left(-\epsilon_{DRD}^{2}k\right)+{\cal O}\left(1/\epsilon_{GC}^{2}k\right)\right].
\]

Recall that $\left(1-f_{I}\right)k=\beta_{c}\left(1-e^{-cP_{S}}\right)k$,
we conclude that, with probability
\[
1-{\cal O}\left[\exp\left(-\epsilon_{DS}^{2}k\right)+\exp\left(-\epsilon_{S}^{2}k\right)+\exp\left(-\epsilon_{D}^{2}k\right)+\exp\left(-\epsilon_{RD}^{2}k\right)+\exp\left(-\epsilon_{DRD}^{2}k\right)+{\cal O}\left(1/\epsilon_{GC}^{2}k\right)\right],
\]

there exists $\epsilon_{I}$ such that

\[
\left(1-\epsilon_{I}\right)f_{I}\leq g_{I}\leq\left(1+\epsilon_{I}\right)f_{I}.
\]

Here, $\epsilon_{I}$ scales as ${\cal O}\left(\epsilon_{DS}+\epsilon_{S}+\epsilon_{D}+\epsilon_{RD}+\epsilon_{DRD}+\epsilon_{GC}\right)$.
Choose all the $\epsilon$'s to be $k^{-1/3}$. Then, $\epsilon_I$ scales as ${\cal O}\left(k^{-1/3}\right)$ with probability $1-{\cal O}\left(k^{-1/3}\right)$.

\subsection{Geometric-decay Phase}

\subsubsection{Stage $l+1$ ($0\leq l\leq L-1$)}
\begin{itemize}
\item The probability that a right node being a resolvable multiton:
\end{itemize}
\begin{eqnarray*}
Q_{M}^{(II,l+1)} & = & {g_{II,l}k \choose 1}\frac{1}{f_{II,l}k}\left(1-\frac{1}{f_{II,l}k}\right)^{g_{II,l}k-1}\left[1-\left(1-\frac{1}{f_{II,l}k}\right)^{\left(1-g_{II,l}\right)k}\right]\\
 & = & \frac{g_{II,l}}{f_{II,l}}\left[\left(1-\frac{1}{f_{II,l}k}\right)^{g_{II,l}k-1}-\left(1-\frac{1}{f_{II,l}k}\right)^{k-1}\right]\\
 & \doteq\frac{g_{II,l}}{f_{II,l}} & \left(e^{-\frac{g_{II,l}}{f_{II,l}}}-e^{-\frac{1}{f_{II,l}}}\right).
\end{eqnarray*}

\begin{itemize}
\item The number of resolvable multitons is bounded by
\[
\left(1+\epsilon_{M}^{(II,l+1)}\right)cf_{II,l}kQ_{M}^{(II,l+1)}
\]
and
\[
\left(1-\epsilon_{M}^{(II,l+1)}\right)cf_{II,l}kQ_{M}^{(II,l+1)}
\]
with probability $1-{\cal O}\left(\exp\left(-\left[\epsilon_{M}^{\left(II,l+1\right)}\right]^{2}f_{II,l}k\right)\right)$ for any $\epsilon_{M}^{\left(II,l+1\right)}>0$.
\item The number of non-zero components which are recovered (both magnitude
and phase)
\end{itemize}
Let $\epsilon_{II,0}=\epsilon_I$. By Lemma \ref{lem:couponcollection}, we know that

\begin{eqnarray*}
g_{II,l+1}-g_{II,l} & \leq & \left(1+\epsilon_{DM}^{(II,l+1)}\right)\left[1-e^{-\left(1+\epsilon_{M}^{(II,l+1)}\right)cf_{II,l}Q_{M}^{(II,l+1)}/g_{II,l}}\right]g_{II,l}\\
 & \doteq & \left(1+\epsilon_{DM}^{(II,l+1)}\right)\left[1-e^{-\left(1+\epsilon_{M}^{(II,l+1)}\right)c\left(e^{-\frac{g_{II,l}}{f_{II,l}}}-e^{-\frac{1}{f_{II,l}}}\right)}\right]g_{II,l}\\
 & \leq & \left(1+\epsilon_{DM}^{(II,l+1)}\right)\left[1-e^{-\left(1+\epsilon_{M}^{(II,l+1)}\right)c\left(e^{-\left(1-\epsilon_{II,l}\right)}-e^{-\frac{1}{f_{II,l}}}\right)}\right]g_{II,l}
\end{eqnarray*}

and

\begin{eqnarray*}
g_{II,l+1}-g_{II,l} & \geq & \left(1-\epsilon_{DM}^{(II,l+1)}\right)\left[1-e^{-\left(1-\epsilon_{M}^{(II,l+1)}\right)cf_{II,l}Q_{M}^{(II,l+1)}/g_{II,l}}\right]g_{II,l}\\
 & \doteq & \left(1-\epsilon_{DM}^{(II,l+1)}\right)\left[1-e^{-\left(1-\epsilon_{M}^{(II,l+1)}\right)c\left(e^{-\frac{g_{II,l}}{f_{II,l}}}-e^{-\frac{1}{f_{II,l}}}\right)}\right]g_{II,l}\\
 & \geq & \left(1-\epsilon_{DM}^{(II,l+1)}\right)\left[1-e^{-\left(1-\epsilon_{M}^{(II,l+1)}\right)c\left(e^{-\left(1-\epsilon_{II,l}\right)}-e^{-\frac{1}{f_{II,l}}}\right)}\right]g_{II,l},
\end{eqnarray*}

with probability $1-{\cal O}\left(\exp\left(-\left[\epsilon_{M}^{\left(II,l+1\right)}\right]^{2}f_{II,l}k\right)+\exp\left(-\left[\epsilon_{DM}^{\left(II,l+1\right)}\right]^{2}g_{II,l}k\right)+k^{-1/3}\right)$
by Lemma \ref{lem:couponcollection} and Union bound for any $\epsilon_{DM}^{\left(II,l+1\right)}>0$.

We conclude there exists $\epsilon'_{II,l+1}$ such that

\begin{eqnarray*}
e^{-cP_{M}^{(II,l+1)}}\left(1-\epsilon'_{II,l+1}\right)g_{II,l} & \leq & g_{II,l+1}\\
 & \leq & e^{-cP_{M}^{(II,l+1)}}\left(1+\epsilon'_{II,l+1}\right)g_{II,l}.
\end{eqnarray*}

Here $\epsilon'_{II,l+1}$ scales as ${\cal O}\left[\epsilon_{DM}^{(II,l+1)}+\epsilon_{M}^{(II,l+1)}+\epsilon_{II,l}\right]$.

Since $\left(1-\epsilon_{II,l}\right)f_{II,l}\leq g_{II,l}=\left(1+\epsilon_{II,l}\right)f_{II,l}$,
we have

\begin{eqnarray*}
e^{-cP_{M}^{(II,l+1)}}\left(1-\epsilon'_{II,l+1}\right)\left(1-\epsilon_{II,l}\right)f_{II,l} & \leq & g_{II,l+1}\\
 & \leq & e^{-cP_{M}^{(II,l+1)}}\left(1+\epsilon'_{II,l+1}\right)\left(1+\epsilon_{II,i}\right)f_{II,l}.
\end{eqnarray*}

Since $f_{II,l+1}=e^{-cP_{M}^{(II,l+1)}}f_{II,l}$, we have

\begin{eqnarray*}
\left(1-\epsilon'_{II,l+1}\right)\left(1-\epsilon_{II,l}\right)f_{II,l+1} & \leq & g_{II,l+1}\\
 & \leq & \left(1+\epsilon'_{II,l+1}\right)\left(1+\epsilon_{II,l}\right)f_{II,l+1}.
\end{eqnarray*}

Thus, we get that

\begin{eqnarray*}
\left(1-\epsilon_{II,l+1}\right)f_{II,l+1} & \leq & g_{II,l+1}\\
 & \leq & \left(1+\epsilon_{II,l+1}\right)f_{II,l+1}.
\end{eqnarray*}

Here, $\epsilon{}_{II,l+1}$ scales as ${\cal O}\left[\epsilon_{DM}^{(II,l+1)}+\epsilon_{M}^{(II,l+1)}+2\epsilon_{II,l}\right]$.

Choose $\epsilon_{DM}^{(II,l+1)}$ and $\epsilon_{M}^{(II,l+1)}$ to be $k^{-1/3}$ for all $l$. The error probability in each stage is ${\cal O}\left(k^{-1/3}\right)$ (which is the dominant term).

After $L$ stages ($L={\cal O}(\log\log k)$),

\begin{equation}
\label{equ:bound}
\left(1-\epsilon_{II,L}\right)f_{II,L}\leq g_{II,L}\leq\left(1+\epsilon_{II,L}\right)f_{II,L}
\end{equation}

holds with probability $1-{\cal O}\left(\log\log k\cdot k^{-1/3}\right)$ by Union bound.
Here $\epsilon_{II,L}$ scales as ${\cal O}\left(\log k\cdot k^{-1/3}\right)$
(since each stage $\epsilon_{II,l}$ doubles).

\subsection{Cleaning-up phase}
\begin{thm}
\label{thm:coupon}{[}Folklore{]}(Coupon Collection) Let the random
variable $X$ denote the minimum number of trials for collecting each
of the $V$ types of coupons. Then, we have $\Pr[X>\eta V\log V]\leq V^{-\eta+1}$ for any $\eta>0$.\end{thm}
\begin{itemize}
\item The probability that a right node being a resolvable multiton:
\begin{eqnarray*}
Q_{M}^{III} & \doteq & \frac{g_{II,L}}{f_{II,L}}\left(e^{-\frac{g_{II,L}}{f_{II,L}}}-e^{-\frac{1}{f_{II,L}}}\right).
\end{eqnarray*}

\item $f_{III}=1/\log k$. The number of resolvable multitons is lower bounded
by
\[
\left(1-\epsilon_{M}^{III}\right)cf_{III}k\log\left(f_{III}k\right)Q_{M}^{III},
\]
 with probability $1-{\cal O}\left(\exp\left(-\left[\epsilon_{M}^{III}\right]^{2}k/\log k\right)\right)$ given Equation (\ref{equ:bound}) holds  for any $\epsilon_{M}^{\left(III\right)}>0$.
\item The number of unresolved components in this phase is ${\cal O}(k/\log k)$.
By Theorem \ref{thm:coupon}, all the components are resolved with
probability $1-{\cal O}(\log k/k)$ for large enough $c$ given Equation (\ref{equ:bound}) holds.
\end{itemize}
By Union bound, the overall error probability is $o(1)$.

\section{Conclusion}
\label{sec:conclude}
In this paper, we present the first algorithm for compressive phase retrieval problem whose number of measurements is order-optimal and computational complexity is nearly order-optimal.

\section{Appendix}
\subsection{Proof for Lemma \ref{lem:couponcollection}}
\begin{IEEEproof}
Let $Y_{i}$ be the indicator random variable which represents whether
$i$-th coupon is picked in $M$ trials. We know that $Y_{i}'s$ are
dependent and

\[
Y_{i}=\begin{cases}
1 & \mbox{with probability \ensuremath{1-\left(1-\frac{1}{V}\right){}^{M}}}\\
0 & \mbox{with probability \ensuremath{\left(1-\frac{1}{V}\right){}^{M}}}.
\end{cases}
\]

Then, $Y=Y_{1}+\cdots+Y_{V}$ is the total number of different types
of coupons picked in $M$ trials.

By the linearity of expectation, we have
\begin{eqnarray*}
E[Y] & = & \Sigma_{i=1}^{V}E[Y_{i}]\\
 & = & V\left[1-\left(1-\frac{1}{V}\right)^{M}\right]\\
 & \doteq & V\left(1-e^{-\frac{M}{V}}\right)\\
 & = & V\left(1-\frac{V-U}{V}\right)\\
 & = & U.
\end{eqnarray*}

Let $Z_{1},\ldots,Z_{M}$ be independent random variables all taking
values in $[V]$ uniformly at random representing each pick for $V$
types of coupon.

Let $f(Z_{1},\ldots,Z_{M})$ be the number of different types of coupons
picked. Then, ${\bf E}[f]={\bf E}[Y]\doteq U$.

Also, $\forall i\in[M]$,

\[
\left|f\left(Z_{1},\ldots,Z_{i},\ldots,Z_{M}\right)-f\left(Z_{1},\ldots,Z'_{i},\ldots,Z_{M}\right)\right|\leq1.
\]
.

For all $\beta>0$, by McDiarmid's Inequality (Theorem \ref{thm:mciarmid}),
we have

\[
\Pr(f-{\bf E}[f]\leq-\beta)\leq\exp\left(-\frac{2\beta^{2}}{M}\right)
\]

and

\[
\Pr(f-{\bf E}[f]\geq\beta)\leq\exp\left(-\frac{2\beta^{2}}{M}\right).
\]

Thus,

\[
\Pr(f\leq V(1-e^{-M/V})-\beta)\leq\exp\left(-\frac{2\beta^{2}}{M}\right).
\]

Let $M=V\log\frac{V}{V-U}$ and $\beta=\epsilon U$, we know that

\begin{eqnarray*}
\Pr(f\leq(1-\epsilon)U) & \leq & \exp\left(-\frac{2(\epsilon U)^{2}}{V\log\frac{V}{V-U}}\right)\\
 & \leq & \exp-\frac{2(\epsilon U)^{2}}{V\frac{U}{V-U}}\\
 & = & \exp\left(-\frac{2\epsilon^{2}U(V-U)}{V}\right)
\end{eqnarray*}

and

\begin{eqnarray*}
\Pr(f\geq(1+\epsilon)U) & \leq & \exp\left(-\frac{2\epsilon^{2}U(V-U)}{V}\right).
\end{eqnarray*}

Therefore,

\begin{eqnarray*}
\Pr((1-\epsilon)U\leq f\leq(1+\epsilon)U) & \geq & 1-2\exp\left(-\frac{2\epsilon^{2}U(V-U)}{V}\right).
\end{eqnarray*}
\end{IEEEproof}
\subsection{Proof of Theorem \ref{thm:coupon}}
\begin{IEEEproof}
Let $Z_{i}$ be the event that $i$-th coupon has not yet picked in
$M$ trials. We know that

\begin{eqnarray*}
\Pr\left(Z_{i}\right) & = & \left(1-\frac{1}{V}\right)^{M}\\
 & \leq & \exp\left(-M/V\right).
\end{eqnarray*}

Then,
\begin{eqnarray*}
\Pr\left(X>M\right) & = & \Pr\left(\cup_{i=1}^{v}Z_{i}\right)\\
 & \leq & \sum_{i=1}^{v}\Pr\left(Z_{i}\right)\\
 & \leq & V\exp\left(-M/V\right).
\end{eqnarray*}

Let $M=\eta V\log V$, we get

\begin{eqnarray*}
\Pr\left(X>\eta V\log V\right) & \leq & V^{-\eta+1}.
\end{eqnarray*}
\end{IEEEproof}
\subsection{Chernoff Bound and McDiamid's Inequality}
\begin{thm}
\label{thm:chernoff}(Chernoff Bound) Let $X_{1},\ldots,X_{n}$ be
independent random variables. Assume that $0\leq X_{i}\leq1$ for
each $i\in[n]$. Let $X=X_{1}+\cdots+X_{n}$. $\mu={\bf E}[X]={\bf E}[X_{1}]+\cdots+{\bf E}[X_{n}]$.
Then for any $\epsilon\geq0$,

\[
\Pr[X\leq(1-\epsilon)\mu]\leq\exp\left(-\frac{\epsilon^{2}}{2}\mu\right)
\]

and

\[
\Pr[X\geq(1+\epsilon)\mu]\leq\exp\left(-\frac{\epsilon^{2}}{2+\epsilon}\mu\right).
\]

\end{thm}
\begin{thm}
\label{thm:mciarmid}\cite{Mcd:89}(McDiarmid's Inequality) Let $X_{1},\ldots,X_{m}$
be independent random variables all taking values in the set $\mathsf{\mathcal{X}}$.
Further, let $f:{\cal X}^{m}\rightarrow\mathbb{R}$ be a function
of $X_{1},\ldots,X_{m}$ that satisfies $\forall i$,$\forall x_{1},\ldots,x_{m}$,
$x_{i}'\in{\cal X}$,
\begin{eqnarray*}
|f(x_{1},\ldots,x_{i},\ldots,x_{m})-f(x_{1},\ldots,x'_{i},\ldots,x_{m})| & \leq & c_{i}.
\end{eqnarray*}

Then for all $\epsilon>0$,

\begin{eqnarray*}
\Pr(f-{\bf E}[f]\geq\epsilon) & \leq & \exp\left(\frac{-2\epsilon^{2}}{\Sigma_{i=1}^{m}c_{i}^{2}}\right)
\end{eqnarray*}

and

\begin{eqnarray*}
\Pr(f-{\bf E}[f]\leq-\epsilon) & \leq & \exp\left(\frac{-2\epsilon^{2}}{\Sigma_{i=1}^{m}c_{i}^{2}}\right).
\end{eqnarray*}
\end{thm}

\bibliographystyle{IEEEtran}
\bibliography{sheng}

\begin{thebibliography}{10}
\providecommand{\url}[1]{#1}
\csname url@samestyle\endcsname
\providecommand{\newblock}{\relax}
\providecommand{\bibinfo}[2]{#2}
\providecommand{\BIBentrySTDinterwordspacing}{\spaceskip=0pt\relax}
\providecommand{\BIBentryALTinterwordstretchfactor}{4}
\providecommand{\BIBentryALTinterwordspacing}{\spaceskip=\fontdimen2\font plus
\BIBentryALTinterwordstretchfactor\fontdimen3\font minus
  \fontdimen4\font\relax}
\providecommand{\BIBforeignlanguage}[2]{{%
\expandafter\ifx\csname l@#1\endcsname\relax
\typeout{** WARNING: IEEEtran.bst: No hyphenation pattern has been}%
\typeout{** loaded for the language `#1'. Using the pattern for}%
\typeout{** the default language instead.}%
\else
\language=\csname l@#1\endcsname
\fi
#2}}
\providecommand{\BIBdecl}{\relax}
\BIBdecl

\bibitem{mil:90}
R.~Millane, ``Phase retrieval in crystallography and optics,'' \emph{J. Opt.
  Soc. Am. A}, vol.~7, pp. 394--411, 1990.

\bibitem{DauF:87}
J.~Dainty and J.~Fienup, ``Phase retrieval and image reconstruction for
  astronomy,'' \emph{In: H. Stark, ed., Image Recovery: Theory and Application,
  Academic Press, New York}, pp. 231--275, 1987.

\bibitem{CRT:06}
E.~Candes, J.~Romberg, and T.~Tao, ``Robust uncertainty principles: exact
  signal reconstruction from highly incomplete frequency information,''
  \emph{IEEE Transactions on Information Theory}, vol.~52, no.~2, pp. 489--509,
  2006.

\bibitem{Fie:82}
J.~Fienup, ``Phase retrieval algorithms: a comparison,'' \emph{Appl. Opt.},
  vol.~21, pp. 2758--2769, 1982.

\bibitem{Fie:13}
------, ``Phase retrieval algorithms: a personal tour [invited],'' \emph{Appl.
  Opt.}, vol.~52, pp. 45--56, 2013.

\bibitem{CPA:13}
E.~Cand\`{e}s, T.~Strohmer, and V.~Voroninski, ``Phaselift: Exact and stable
  signal recovery from magnitude measurements via convex programming,''
  \emph{Communications on Pure and Applied Mathematics}, vol.~66, no.~8, pp.
  1241--1274, 2013.

\bibitem{Waldm:13}
I.~Waldspurger, A.~d'Aspremont, and S.~Mallat, ``Phase recovery, maxcut and
  complex semidefinite programming,'' \emph{Mathematical Programming}, pp.
  1--35, 2013.

\bibitem{CanR:2012}
E.~Cand\`{e}s and B.~Recht, ``Exact matrix completion via convex
  optimization,'' \emph{Commun. ACM}, vol.~55, no.~6, pp. 111--119, Jun. 2012.

\bibitem{CanL:13}
E.~Cand\`{e}s and X.~Li, ``Solving quadratic equations via phaselift when there
  are about as many equations as unknowns,'' \emph{Foundations of Computational
  Mathematics}, pp. 1--10, 2013.

\bibitem{AleBFM:12}
B.~Alexeev, A.~S. Bandeira, M.~Fickus, and D.~G. Mixon, ``Phase retrieval with
  polarization,'' \emph{e-prints, arXiv:1210.7752[cs.IT]}, 2012.

\bibitem{NetJS:13}
P.~Netrapalli, P.~Jain, and S.~Sanghavi, ``Phase retrieval using alternating
  minimization,'' \emph{e-prints, arXiv:1306.0160v1 [stat.ML]}, 2013.

\bibitem{HenY:13}
H.~Ohlsson and Y.~C. Eldar, ``On conditions for uniqueness in sparse phase
  retrieval,'' \emph{e-prints, arXiv:1308.5447[cs.IT]}, 2013.

\bibitem{MehV:13}
M.~Ak\c{c}akaya and V.~Tarokh, ``New conditions for sparse phase retrieval,''
  \emph{e-prints, arXiv:1310.1351[cs.IT]}, 2013.

\bibitem{OhlYDS:11}
H.~Ohlsson, A.~Yang, R.~Dong, and S.~Sastry, ``Compressive phase retrieval from
  squared output measurements via semidenite programming,'' \emph{e-prints,
  arXiv:1111.6323v3 [math.ST]}, 2011.

\bibitem{LiV:12}
X.~Li and V.~Voroninski, ``Sparse signal recovery from quadratic measurements
  via convex programming,'' \emph{e-prints, arXiv:1209.4785[cs.IT]}, 2012.

\bibitem{SheBE:13}
Y.~Shechtman, A.~Beck, and Y.~C. Eldar, ``Gespar: Effcient phase retrieval of
  sparse signals,'' \emph{e-prints, arXiv:1301.1018[cs.IT]}, 2013.

\bibitem{JanOH:12}
K.~Jaganathan, S.~Oymak, and B.~Hassibi, ``Recovery of sparse 1-d signals from
  the magnitudes of their fourier transform,'' in \emph{2012 IEEE International
  Symposium on Information Theory Proceedings (ISIT)}, 2012, pp. 1473--1477.

\bibitem{JanOH:13}
------, ``Sparse phase retrieval: Convex algorithms and limitations,'' in
  \emph{2013 IEEE International Symposium on Information Theory Proceedings
  (ISIT)}, 2013, pp. 1022--1026.

\bibitem{JanOH:121}
------, ``Phase retrieval for sparse signals using rank minimization,'' in
  \emph{2012 IEEE International Conference on Acoustics, Speech and Signal
  Processing (ICASSP)}, 2012, pp. 3449--3452.

\bibitem{SamAMYB:13}
S.~Oymak, A.~Jalali, M.~Fazel, Y.~C. Eldar, and B.~Hassibi, ``Simultaneously
  structured models with application to sparse and low-rank matrices,''
  \emph{e-prints, arXiv:1212.3753[cs.IT]}, 2013.

\bibitem{PawR:12}
S.~Pawar and K.~Ramchandran, ``A hybrid dft-ldpc framework for fast, efficient
  and robust compressive sensing,'' in \emph{2012 50th Annual Allerton
  Conference on Communication, Control, and Computing (Allerton)}, 2012, pp.
  1943--1950.

\bibitem{SHO-FA}
M.~Bakshi, S.~Jaggi, S.~Cai, and M.~Chen, ``Sho-fa: Robust compressive sensing
  with order-optimal complexity, measurements, and bits,'' in \emph{2012 50th
  Annual Allerton Conference on Communication, Control, and Computing
  (Allerton)}, 2012, pp. 786--793.

\bibitem{Bol:01}
B.~Bollobas, ``Random graphs,'' \emph{Cambridge University Press}, 2001.

\bibitem{MotR:95}
R.~Motwani and P.~Raghavan, ``Randomized algorithms,'' \emph{Cambridge
  University Press}, 1995.

\bibitem{Tar:72}
R.~Tarjan, ``Depth first search and linear graph algorithms,'' \emph{SIAM
  Journal on Computing}, 1972.

\bibitem{Mcd:89}
C.~McDiarmid, ``On the method of bounded differences,'' \emph{Surveys in
  combinatorics}, vol. 141, no.~1, pp. 148--188, 1989.

\end{thebibliography}
\end{document}